\documentstyle[sprocl,epsfig]{article}

\newcommand{\al}{\alpha}
\newcommand{\be}{\beta}

\def\Journal#1#2#3#4{{#1} {\bf #2}, #3 (#4)}

\def\NPB{{\em Nucl. Phys.} B}
\def\PLB{{\em Phys. Lett.}  B}
\def\PRL{\em Phys. Rev. Lett.}
\def\PRD{{\em Phys. Rev.} D}
\def\EPC{{\em Eur. Phys. J.} C }
\def\ZPC{{\em Z. Phys.} C}
\def\PTP{\em Prog.~Theor.~Phys.}

\def\al{\alpha}

\def\bea{\begin{eqnarray}}
\def\eea{\end{eqnarray}}

\begin{document}

\title{Summary Report for Electroweak Symmetry Breaking Session in 
       LCWS 2004}

\author{SHINYA KANEMURA}

\address{Department of Physics, Osaka University, Toyonaka, Osaka
         560-0043, Japan}


\maketitle\abstracts{
Theoretical activities on the Higgs physics and its  
implication to physics beyond the standard model 
are summarized as the summary report for the Higgs 
session in LCWS 2004. }


\section{Introduction}

The origin of electroweak symmetry breaking has been of 
a central interest in high energy physics for over two decades, 
and will continue being so in future until it will 
be clarified by the experiment. 
In the standard model (SM), 
the electroweak symmetry is spontaneously broken by 
introducing an iso-doublet scalar field, the Higgs field. 
Its neutral component receives the vacuum expectation value ($v$).  
The weak gauge bosons then obtain their masses 
through the Higgs mechanism.
At the same time, all quarks and charged leptons receive the 
masses from the Yukawa interactions with the Higgs field. 
Moreover, the Higgs boson ($h$) itself is 
also given its mass ($m_h$) by the vacuum expectation value 
through the self-interaction of the Higgs boson. 
All these masses of the SM particles 
are expressed as multiplication of coupling constants 
with $v$. 
Therefore, 
there is an universality 
between masses and coupling constants:  
\begin{eqnarray}
\frac{2 m_W}{g} = \frac{\sqrt{2} m_t}{y_t} 
= \frac{\sqrt{2} m_b}{y_b} 
= \frac{\sqrt{2} m_{\tau}}{y_\tau} = ... = \frac{ m_h}{\sqrt{2 \lambda}} 
= v \simeq 246 {\rm GeV}, 
\end{eqnarray}
where $g$ is the weak gauge coupling, $y_f$ is the Yukawa coupling 
constant to the fermion $f$, $\lambda$ is the self-coupling 
constant of $h$, and $m_i$ is the mass of the field $i$. 
The SM can be tested through this universality. 

The Higgs boson $h$ is yet to be discovered, 
and its mass $m_h$ remains unknown. 
In the SM, $m_h$ is a parameter which 
characterizes the property of the Higgs dynamics. 
Since $m_h \propto \sqrt{\lambda} v$, 
a light $h$ means that the Higgs sector is weakly-interacted, 
while a heavy $h$ corresponds to the strong coupling.
Although $m_h$ is an unknown parameter, 
by putting the requirement that the theory must be consistent 
up to a given value of the cutoff scale $\Lambda$,  
the allowed region of $m_h$ can be predicted 
as a function of $\Lambda$~\cite{lindner}.  
Based on this requirement, a renormalization group equation 
analysis for the coupling constant $\lambda$ gives  
upper and the lower bounds on $m_h$ for a given $\Lambda$. 
In the SM, for $\Lambda=10^{19}$ GeV, the allowed region of $m_h$ 
is evaluated as about $135 < m_h < 180$ GeV 
while for $\Lambda=10^{3}$ GeV it is about $m_h < 480$ GeV.
Hence, as long as the SM Higgs sector is assumed, 
a light Higgs boson would indicate 
a weakly coupled theory with a high cut-off scale. 
Scenarios based on grand unified theories (GUTs) 
might correspond to this case. 
In such cases, supersymmetry (SUSY) would necessarily be required 
to reduce the problem of 
large hierarchy between the weak scale and the scale of $\Lambda$.
On the contrary, a heavy Higgs boson with $m_h \sim$ several hundred GeV 
would imply a strongly-coupled Higgs sector with a low 
$\Lambda \sim {\cal O}(1)$ TeV.
In such a case, the Higgs sector should be considered 
as an effective theory of a new dynamics at TeV scales. 
Therefore, from knowing the mass of the Higgs boson, a useful hint 
for new physics beyond the SM can be obtained.

In addition to the mass of the Higgs boson, tri-linear couplings of  
$hVV$ ($g_{hVV}^{} g_{\mu\nu}$), where $VV=ZZ$ and $W^+W^-$, 
and $hhh$ ($\lambda_{hhh}$) are particularly important 
to test the Higgs sector.   
These couplings are absent in the Lagrangian, 
and are induced according to electroweak symmetry breaking 
from the $hhVV$ and $hhhh$ interactions, respectively.
In the SM, the $hZZ$ and $hhh$ couplings can be expressed 
in terms of $m_Z^{}$ and $m_h$ as~\cite{kkosy}: 
\begin{eqnarray}
   g_{hZZ}^{} = \frac{2 m_Z^2}{v}  
   \left( 1- \frac{5}{32\pi^2}  \frac{m_t^2}{v^2} + ... \right), 
\;\;
  \lambda_{hhh} = - \frac{3 m_h^2}{v} 
   \left( 1 - \frac{1}{\pi^2} \frac{m_t^4}{v^2 m_h^2} + ...\right), 
\label{eq2}
\end{eqnarray}
The first equation is useful to directly test the nature 
of the Higgs mechanism, while the second one can be 
used to explore the structure of Higgs sectors.
Moreover, precise determination of the Yukawa coupling constants 
is essential to study the structure of the fermion mass generation.


Experimental identification of the Higgs boson is 
one of the most important goals of high energy collider experiments. 
The LEP Electroweak Working Group fit favors a relatively 
light Higgs boson with its mass below 251 GeV, 
assuming the SM~\cite{LEPEWWG}.
The search for the Higgs bosons is being carried at Fermilab
Tevatron and will be continued at CERN Large Hadron Collider (LHC).
There, the SM Higgs boson is expected to be discovered 
as long as its mass is less than 1 TeV. 
An electron-positron linear collider (LC) 
and its photon collider option can provide good opportunity 
for precise measurement of the Higgs boson couplings. 
At a LC, the Higgs boson $h$ is produced mainly via the 
Higgsstrahlung process $e^+e^- \to Z h$
for relatively low energies and also 
via the fusion process   
$e^+e^- \to W^{+\ast}W^{-\ast} \nu \bar\nu \to h \nu \bar\nu$ 
for higher energies~\cite{eehx}. 
In both production mechanisms, the Higgs boson is produced 
through the coupling with weak gauge bosons $hVV$.  
The cross sections are expected to be measured at a percent 
level or better unless the Higgs boson is relatively heavy. 
The Higgs boson couplings with heavy quarks (except the top quark) 
and the tau lepton 
can be tested by measuring the decay branching 
ratios of the Higgs boson. 
Furthermore, the tri-linear Higgs coupling 
$hhh$~\cite{battaglia,yasui,yamashita} 
and the top-Yukawa coupling $h t \bar t$
can be determined by 10-20\% accuracy by measuring the cross section 
of double Higgs production processes $e^+e^- \to Z h h$ as well as  
$e^+e^- \to W^{+\ast}W^{-\ast} \nu \bar\nu \to h h \nu \bar\nu$ 
and the top-associated Higgs production process~\cite{htt}
$e^+e^- \to  h t \bar t$, respectively. 
The $\gamma\gamma$ option of the LC can also be useful 
for determination of the CP property 
and the couplings of the Higgs boson~\cite{gammagamma}.     

\section{Extended Higgs sectors and new physics}

Studying the Higgs sector is not only useful for the 
confirmation of the breaking mechanism of the electroweak gauge
symmetry in the SM, but also provides a sensitive window for new physics 
beyond the SM. In many models of new physics, an extended 
Higgs sector from the SM one appears in the low energy effective theory, 
which has discriminative phenomenological properties. 
In extended Higgs sectors, there is mixing among  
more than one Higgs bosons, so that 
the relations in Eq.~(\ref{eq2}) are modified 
due to the mixing effect~\cite{dec-region,hhh_ext} as well as 
the loop effect of extra Higgs bosons~\cite{hhh_hollik,kkosy}. 
If an extended Higgs sector such as the two Higgs doublet model 
(THDM) or some other 
extension of the SM Higgs sector is assumed, 
the theoretical bounds on the mass of the lightest Higgs boson 
are changed for a given cut-off scale~\cite{kko,zee}. 

One popular example for new physics models 
is the minimal supersymmetric standard 
model (MSSM), in which the Higgs sector is a THDM~\cite{HHG}. 
The most important prediction in the MSSM is that for the mass 
of the lightest Higgs boson~\cite{heinemeyer}. 
The coupling constants in the Higgs potential 
are given by the electroweak gauge coupling constants 
which are originated from the D-term contribution.
Thus, the mass of the lightest CP-even Higgs boson 
is limited at tree level to be less than the mass of the $Z$ boson, 
and at loop levels the upper limit is shifted to be around 
135 GeV by radiative corrections~\cite{mssmhiggs}.  
The predicted mass $m_h$ of the Higgs boson is sensible to 
the SUSY parameters, so that both precise measurement of $m_h$
and unambiguous theoretical prediction for $m_h$ are quite important 
to obtain the detailed information of SUSY~\cite{heinemeyer}.  

Some models based on dynamical breaking of the electroweak symmetry
also require more than one Higgs doublets in the low energy 
effective theories~\cite{TC}. 
In general such models induce relatively strong 
coupling constants in the effective Higgs potential, 
and the typical masses of the Higgs bosons are 
larger than the lightest Higgs boson in weakly coupled models. 
As discussed, the present experimental data~\cite{LEPEWWG} indicate 
a weakly interacting Higgs boson in the minimal Higgs sector.  
This experimental upper limit, however, strongly depends on the model. 
In extended Higgs models, there are additional 
contributions to the weak boson two-point functions which could 
cancel those of the SM Higgs boson loop, so that 
the lightest Higgs boson heavier than 251 GeV can be 
allowed~\cite{peskin-wells}. 
Hence, in order for strongly-coupled dynamical models  
to be consistent with the low energy data, an extended Higgs 
sector is required in the low energy effective theory. 

There are other motivations to introduce extra Higgs fields 
at low energy, such as electroweak baryogenesis~\cite{ewbg}, 
neutrino mass problem~\cite{zee}, and 
top-bottom mass hierarchy~\cite{top-bottom}.  


A common feature of extended Higgs sectors is the existence 
of additional scalar bosons, such as charged Higgs bosons 
and CP-odd Higgs boson(s). 
Discovery of these extra scalar particles would
directly show a new physics model beyond   
the SM.  
Once these additional Higgs bosons are found, 
we would be able to test each new physics model 
by measuring details of their properties.
Various production processes of heavy neutral Higgs 
bosons and the charged Higgs boson at a LC are studied
in the MSSM (THDM)~\cite{LC-heavyHiggs1}. 
A $\gamma\gamma$ collider provides further opportunity 
for the heavy Higgs search, where various properties 
of the heavy Higgs bosons, such as CP-property, 
can be explored~\cite{gammagamma,maria,asakawa,LC-heavyHiggs2}.


Even if the extra Higgs bosons are not found, 
we can still obtain insight by looking for indirect 
effects of the extra Higgs boson from the precise  
determination of the lightest Higgs boson properties~\cite{LC-indirect}. 
General features of such indirect effects can be learned 
through the study of decoupling property in the THDM~\cite{dec-region}. 
The tree-level couplings with weak gauge bosons $hVV$ ($VV=WW$ and $ZZ$) 
are changed by the factor of $\sin(\beta-\alpha)$, where $\alpha$ is the 
mixing angle of CP-even Higgs bosons and $\tan\beta$ is  
the ratio of the vacuum expectation values.     
The other couplings such as $h f \bar f $ ($f$: a fermion) 
and $hhh$ are also changed. In the limit $\sin(\be-\al) \to 1$, 
these couplings become the same as those in the SM. 
In addition to the mixing effect, the couplings can 
receive large quantum corrections provided that 
the extra Higgs bosons have the non-decoupling property. 
When their masses are predominantly generated by $v$, contributions 
in powers of the mass of the loop particles 
can appear in the one loop effect on the couplings of 
$hZZ$ and $hhh$, similar to the top-quark loop effect seen 
in Eq.~(\ref{eq2}).  
They are quadratic for the $hVV$ coupling and quartic 
for the $hhh$ coupling~\cite{kkosy,okada}. 
Similar non-decoupling effects~\cite{nondec_THDM} can also appear in 
$h \to \gamma\gamma$~\cite{hgammagamma,hbbbar}, 
$h \to b \bar b$~\cite{hbbbar}, 
$e^+e^- \to W^+W^-$~\cite{eeww} and those with 
the coupling $W^\pm H^\mp Z$~\cite{whz}.
These observables can receive significant corrections 
due to the non-decoupling effect even in the SM-like limit 
$\sin(\beta-\alpha) \to 1$. 
On the contrary, when the heavy Higgs bosons obtain the masses 
mainly from an invariant mass parameter, such power-like 
contribution disappears, and the loop effects vanish 
in the limit where the mass of additional Higgs bosons 
is large. 
The MSSM Higgs sector belongs to this case. 
A systematic approach to the correlation in the deviations among 
various observables is necessary for such cases~\cite{LC-indirect}. 


\section{Activities in Electroweak Symmetry Breaking Session}

In the following, I summarize some of the theoretical 
contributions to the Higgs session in LCWS 2004. 
Those related to photon colliders and experimental topics 
are treated elsewhere~\cite{gammagamma,barklow}.  


The ability of precise measurement at a future LC 
can be useful for determination of coupling constants,   
only when ambiguity in theory prediction is 
sufficiently suppressed. 
In particular, the production cross sections 
and decay branching ratios of the Higgs boson 
have to be evaluated as precise as possible with 
including higher order contributions of perturbation 
especially in the bench mark theory such as the SM and the MSSM. 
%
%
In the SM, 
the results of complete electroweak $\mathcal{O}(\alpha)$ radiative 
corrections to the Higgs boson production cross sections 
$e^+e^- \to \nu_{\ell} \bar \nu_{\ell} h$ and 
$e^+e^- \to t \bar t h$  were presented
by Dittmaier~\cite{dittmaier}.  
%
%
In the MSSM, the mass of the lightest Higgs boson $m_h$ 
is an output quantity, which is sensible to the SUSY parameters. 
The precise determination of $m_h$ at experiment is quite important 
to obtain the detailed information of SUSY.  
At a LC, the expected experimental error of the $m_h$ measurement 
can be 50 MeV. 
Heinemeyer claimed that the intrinsic error in the theoretical  
prediction of $m_h$ can be reduced to 0.5-0.1 GeV by   
a full two loop calculation with including the leading 
three loop leading contribution~\cite{heinemeyer}.   
The one-loop corrected decay rates 
of the heavy neutral Higgs bosons 
into a pair of charginos and neutralinos 
are evaluated in the MSSM by Eberl~\cite{eberl}.  


In recent years, the MSSM with additional CP phases  
(CPVMSSM) has been studied intensively by many authors~\cite{cpvmssm}.  
The phenomenology of the CPVMSSM is shown to be 
largely different from the CP conserving 
MSSM due to the effect of CP violating phases. 
Akeroyd discussed the case in which the typical SUSY 
breaking scale ($M_{SUSY}^{}$) is as large as several TeV~\cite{akeroyd}. 
With such values of $M_{SUSY}^{}$, the predictions on 
the EDM processes are suppressed below the current 
experimental upper limits. 
The phenomenology of CP violation in the non-SUSY THDM 
was also discussed~\cite{maria,cpvthdm,kylee}. 

%
%
The simplest extension of the MSSM is known as 
the next-to-MSSM~\cite{HHG}. Miller studied 
details of the phenomenology in a specific scenario 
in which the lightest SUSY particle (LSP) is the 
singlino, the super-partner of the additional singlet 
field~\cite{djmiller}.  
The phenomenology of the next-to-MSSM with CP violating phases 
was also discussed by Gunion~\cite{nmssmgunion}.

%
%


Baryogenesis is one of the fundamental cosmological 
problems. 
Electroweak baryogenesis~\cite{ewbg} is a scenario 
of baryogenesis at the electroweak phase transition.  
The Higgs sector plays an essentially important role.  
For electroweak baryogenesis it is required 
that the phase transition is strongly first order. 
Okada discussed the phenomenological implication 
of a viable scenario in the THDM~\cite{okada}. 
It was shown that the condition for successful 
baryogenesis on the finite temperature effective potential 
exactly corresponds to the condition of large non-decoupling 
loop effect on the $hhh$ coupling due to 
extra Higgs boson loops for the lightest Higgs boson. 
Such large non-decoupling effects deviates the $hhh$ coupling 
by 10\% or more~\cite{kkosy}, so that the scenario of 
electroweak baryogenesis can be tested by measuring the 
$hhh$ coupling at a LC.


Extra dimensions could provide a candidate for new physics 
which removes the hierarchy problem. 
In models of large extra dimensions, there is an interaction 
between the Ricci scalar curvature and the Higgs doublet field. 
A phenomenological consequence can be an invisible decay 
of the Higgs boson to Kaluza Klein graviscalars. 
Dominici~\cite{dominici}
discussed that the corresponding invisible width 
can cause a suppression of the LHC rates of a light Higgs 
in the visible channels below 5 $\sigma$ for some case.  
In such a case the Higgs boson can be discovered through 
its invisible decay. The combination of the 
measurements at the LHC and the LC can determine 
the parameters of the model.



Lepton flavor violation (LFV) in charged leptons 
directly indicates new physics beyond the SM.  
In the models such as based on SUSY, in addition to
the gauge boson mediated LFV process, the LFV Yukawa 
couplings are naturally induced by slepton mixing~\cite{lfvyukawa}.  
The direct search of such LFV Yukawa coupling 
would be possible by measuring the Higgs decay 
process of $h^0 \to \tau^\pm \mu^\mp$ at a LC. 
Ota discussed details of this possibility, and 
numerically showed the feasibility of this process
from the Higgsstrahlung process at a LC, 
under the constraint from current data of LFV tau 
decay processes~\cite{ota}.

\section{Conclusions}

To explore the nature of electroweak symmetry breaking 
with its implication to new physics is top priority in high energy  
physics. 
The precision measurements of the Higgs sector 
are attained at a LC and its photon collider option, 
at which details of the model can be studied. 
In order to compare theoretical predictions with 
precision data at a LC, highly precise calculations 
for the Higgs boson observables are needed 
in the bench mark theories such as the SM and the MSSM.  
In addition, there are many new physics models beyond 
the SM or the MSSM, which would induce various discriminative 
extended Higgs sectors in the low energy effective theory. 
Phenomenological features of these extended Higgs sectors 
should be more studied to explore new physics.
There remain many things for theorists 
to do for the Higgs physics at a LC.


This work was supported in part by a Grant-in-Aid of the Ministry of 
Education, Culture, Sports, Science and Technology, Government of Japan, 
No. 13047101.

\end{document}